\documentclass[a4paper,11pt]{article}

\usepackage[
linesnumbered,ruled,vlined]{algorithm2e}

\usepackage{a4wide}
\usepackage{graphicx} 
\usepackage[utf8]{inputenc}
\usepackage[english]{babel}
\usepackage[T1]{fontenc}  
\usepackage{amsmath}
\usepackage{amssymb}
\usepackage{algpseudocode}
\usepackage{algorithmicx}
\usepackage{algcompatible}
\usepackage{caption}
\usepackage{csquotes}
\usepackage{subcaption}
\usepackage{hyperref} 
\usepackage{tabularx}
\usepackage{multirow}
\usepackage{comment}
\usepackage{balance}

\newcommand{\FPFC}[1]{\includegraphics[width=0.3\linewidth]{plots/boxplots_#1.pdf}} 
 
\newcommand{\cmmnt}[1]{} 

\newcommand\acksname{Acknowledgments}
\specialcomment{acks}{%
  \begingroup
  \section*{\acksname}
  \phantomsection\addcontentsline{toc}{section}{\acksname}
}

\title{Dynamic Test Case Prioritization in Industrial Test Result Datasets}
\author{
Alina Torbunova \\
Åbo Akademi University \\
Turku, Finland 
\and
Per Erik Strandberg \\
Westermo Network Technologies AB \\
Västerås, Sweden
\and
Ivan Porres \\
Åbo Akademi University \\
Turku, Finland 
}
\date{}

\begin{document}

\maketitle

\begin{abstract}
Regression testing in software development checks if new software features affect existing ones. Regression testing is a key task in continuous development and integration, where software is built in small increments and new features are integrated as soon as possible. It is therefore important that developers are notified about possible faults quickly. In this article, we propose a test case prioritization schema that combines the use of a static and a dynamic prioritization algorithm.  The dynamic prioritization algorithm rearranges the order of execution of tests on the fly, while the tests are being executed. We propose to use a conditional probability dynamic algorithm for this. We evaluate our solution on three industrial datasets and utilize Average Percentage of Fault Detection for that\cmmnt{, as well as our own one which we called Misplaced Tests Score}.\cmmnt{In addition to this, we compare our results to other studies.} The main findings are that our dynamic prioritization algorithm can:
a) be applied with any static algorithm that assigns a priority score to each test case
b) can improve the performance of the static algorithm if there are failure correlations between test cases
c) can also reduce the performance of the static algorithm, but only when the static scheduling is performed at a near optimal level.

Keywords\textemdash regression testing; Test Case Prioritization (TCP); dynamic prioritization
\end{abstract}

\section{Introduction}
Regression testing is used in Continuous Integration (CI) \cite{PanBGB22slr} to ensure that new code changes do not cause problems with existing functionality \cite{LimaV20coleman}. Since each test suite may require an ostensibly long time to run, it is desirable that failing tests are executed as early as possible so that developers can be notified quickly. This means that for each CI cycle (which is an iteration performed when the software is modified), there should be a technique that prioritizes test cases that are most likely to fail.

The techniques applied to regression testing can be divided into three different types: test suite minimization, test case selection, and test case prioritization \cite{YooH2012minimization}.\cmmnt{The first technique is test suite minimization which aims to minimize the size of a test suite by removing unneeded tests which cover the already satisfied criteria. The second one, test case selection, also minimizes the number of test cases, but in this case the focus is on choosing test cases that are relevant for the current CI cycle. The last technique is called test case prioritization and, unlike the first two ones, is concentrated not on selecting a subset of test cases, but on ordering of these test cases. This technique aims to order tests in such a sequence that will provide as high performance as possible with respect to a chosen criterion. \cite{YooH2012minimization}} We focus on test case prioritization and will refer to this technique as TCP. To prioritize test cases, feature sets such as code complexity, textual data, coverage information, user input, and history can be used \cite{PanBGB22slr}.\cmmnt{Four former feature sets may either be unavailable for a prioritization technique or require extra time to extract these features, which can be crucial in terms of time consumed.} In our approach, we use historical results from execution of test cases. The techniques to prioritize tests based on historical data can be divided into static and dynamic tests \cite{PradhanWAYL18remap}. The former create a fixed or, in other words, a static schedule which is then executed \cite{PradhanWAYL18remap}, and the results of the current CI cycle are considered only for the next. The latter adjust the order of test cases during execution \cite{PradhanWAYL18remap}, thus verdicts of the already executed test cases are utilized to rearrange the pending ones. For our approach, we focus on the ability of test cases to reveal faults. By fault we mean some defect in a software which is revealed by the test case that failed. We utilize verdicts of test cases and pairwise probability of tests to fail or pass together. Ideally, we want to achieve an optimal schedule where all failing tests are scheduled before the passing ones.

Our solution is based on conditional probability, and we evaluate our algorithm on three industrial datasets. In this paper, we propose an approach that can be applied with any static scheduler that assigns a priority score to each test case. Our approach calculates the conditional probability of failure or success for correlated test cases and is focused on adjusting or rearranging the schedule of test cases based on the results of the current CI cycle. The goal of our study is to improve a schedule created by a static algorithm.\cmmnt{We also compare our results to other studies. In addition to this, we propose to utilize Misplaced Tests Score (MTS) to evaluate performance of prioritization algorithms.}\cmmnt{We answer the following \textbf{research question}: \textit{Does dynamic test case prioritization using conditional test failure probability improve the schedules created by a static test case prioritization algorithm?}} The main contributions of this paper are: a) a schema for test case prioritization using a static and a dynamic TCP algorithms\footnote{The code is available at https://gitlab.abo.fi/stc/dynamic-tcp.} b) a dynamic TCP algorithm based on the conditional probability of failure c) the evaluation of the dynamic conditional probability algorithm in three industrial datasets.\cmmnt{d) the MTS metric, a normalized metric that can be used to evaluate the performance of TCP algorithms and that avoids the pitfalls of APFD, the most common metric used in the research literature.}

\cmmnt{a schema for test case prioritization using two algorithms: a static TCP algorithm that uses information available before the start of the execution of a test suite and a dynamic TCP algorihtm that uses information collected during the execution of a test suite.\footnote{The code is available at https://gitlab.abo.fi/stc/dynamic-tcp}
\item A dynamic TCP algorithm based on the conditional probability of failure.
\item The evaluation of the dynamic conditional probability algorithm in three industrial datasets. We show how this algorithm increases the performance of the static scheduler in most cases.
\item The MTS metric, a normalized metric that can be used to evaluate the performance of TCP algorithms and that avoids the pitfalls of APFD, the most common metric used in the research literature.}

This article is organized as follows. Section~\ref{sec-previous-work} describes previous work in the field. Section~\ref{sec-static-and-dynamic-tcp} provides a general overview of dynamic prioritization, along with our approach, while Section~\ref{sec-dynamic-with-condit-prob} provides a more detailed description of our solution. \cmmnt{Section~\ref{measuring-effectiveness-of-tcp} presents metrics that we use for evaluation.} Sections~\ref{sec-evaluation} and~\ref{sec-conclusions} present the evaluation results and discussion along with the conclusions in corresponding ways.

\section{Previous Work on  Test Case Prioritization}
\label{sec-previous-work}
There is a large variety of techniques that can be applied to prioritize test cases. Among the static ones, some possible approaches are machine learning \cite{PanBGB22slr}, probabilistic inference \cite{MirarabT07bn, MirarabT08bn}, simulated annealing \cite{ZhangDWZWL22sa}, analytical hierarchy process \cite{NayakKT22ahp}, metaheuristic search techniques such as Hill Climbing and genetic algorithms, as well as greedy algorithms \cite{LiHH07searchalgorithms}.

Besides static prioritization, there are dynamic techniques that reprioritize tests during the execution. These techniques apply features and metrics such as textual data, similarity of test cases, and test coverage \cite{QuNXZ07tcpforblaxbox, YuFMRPC19terminator, MarkiegiAES21productlines}. For our approach, the most relevant ones are test verdicts and relations between test cases. One of the proposed techniques is called AFSAC \cite{ChoKL16afsac}. \cmmnt{which consists of a static prioritizer that orders test cases based on a given weight and a dynamic one where correlated tests are reprioritized. The weight is given as a result of statistical analysis of the failure history of a given test case. Correlations between test cases are computed before the execution starts, and the main idea is that tests are counted as correlated if the verdicts of these change to the opposite ones in the next software revision. This information is utilized during execution, because in the event of failure of some test, the remaining tests which are correlated to this one will be assigned a higher priority.}The main differences of our approach are that by correlated tests we assume the tests that failed or passed together in the past, and we apply conditional probability to count a value of this correlation. The other approach is based on the failure history data (FHD) prioritization technique \cite{KimEL17fhd} and has similarities to AFSAC. \cmmnt{which, similarly to AFSAC \cite{ChoKL16afsac}, utilizes statistical analysis of historical failures for static prioritization, and correlations between test cases are applied in the dynamic part to reorder pending tests. Unlike AFSAC, the FHD-prioritization technique applies an additional objective for static prioritization: a test case can test either a single method or several methods.} Unlike \cite{KimEL17fhd}, we do not utilize information concerning the inner structure of a test case. Another example is an approach called REMAP \cite{PradhanWAYL18remap, PradhanWAYL19multiobjectivesearch}.\cmmnt{This approach consists of two static components (mining of relations between test cases based on their execution history as well as static prioritization based on the chosen objectives) and the dynamic one (reordering of test cases based on results of the executed ones). To mine relation rules, the execution history is analyzed to find if failure or success of one or more test cases can cause a failure or success of some other test case respectively. Then, a static schedule is created based on failure rate of each test case and a score defining how many other tests' results the current one can predict. In addition to that, in \cite{PradhanWAYL19multiobjectivesearch} the researchers define execution time of test cases as the third objective for the static prioritizer. In the dynamic part, the main idea is that after execution of the chosen test case, based on its verdict, related tests are moved either to the beginning or to the end of the queue \cite{PradhanWAYL18remap, PradhanWAYL19multiobjectivesearch}.} The difference with REMAP is that our dynamic algorithm is seen as a separate approach, without a certain static one. Additionally, when we analyze the history of test cases, we apply conditional probability to compute relations between these test cases in terms of failure and success and then utilize this information during execution to adjust individual scores of the pending correlated test cases. Another example is an approach named CoDynaQ \cite{ZhuSR18cp}.\cmmnt{where they apply conditional probability to dynamically reprioritize test cases in multi-request environment. This means that new requests with tests are constantly added for execution. In total, the authors implement three different approaches which differ by the arrangement of the queue with the pending test cases that wait to be executed.} Our approach differs from CoDynaQ by the nature of datasets, because we prioritize a given set of tests that should be executed in a given software revision, and new sets of tests are not constantly added. Furthermore, we apply conditional probability in a different way, since we do not utilize the past co-failure distribution of test cases to change scores of the related pending tests.

\cmmnt{In addition to execution results and relations between test cases, there are techniques where such features and metrics as textual data, similarity of test cases, and test coverage are applied for dynamic prioritization. Qu et al.\ \cite{QuNXZ07tcpforblaxbox}, besides execution results utilized by a static prioritizer, apply relations between test cases that reveal the same fault to dynamically rearrange test cases by escalating or deescalating the priorities of these. Yu et al.\ \cite{YuFMRPC19terminator} apply their TCP approach called TERMINATOR to automated User Interface testing and besides historical test results and execution time of test cases, utilize descriptions of test cases to build a Support Vector Machine which is constantly updated based on the results of the current run to reprioritize test cases. Markiegi et al.\ \cite{MarkiegiAES21productlines} apply TCP for product line engineering. Each test case is paired with a certain product, and in this approach each test case can reveal several faults, which is utilized to rearrange test cases, so that tests that may discover unrevealed faults will be prioritized.}

\section{Combined Static and Dynamic Test Prioritization }
\label{sec-static-and-dynamic-tcp}
In this section, we describe a general framework for combined static and dynamic test prioritization. Our problem setting is regression testing in CI development process. We assume that a software under test is developed in small incremental iterations and a CI server builds and tests the system under development automatically as soon as there is a new revision or at certain time intervals, for example every night. The goal is to schedule the tests to be executed according to a given criterion. This criterion is usually that tests that fail should be executed first.

We can formalize our framework as follows. A sequence of software revisions can be defined as $(S_1,T_1) \ldots,(S_N, T_N)$, where $S_i$ is a software revision build and $T_i$ is its associated test suite as a set of tests. Each test in a test suite can have a verdict $p$ (pass) or $f$ (fail). A schedule $S$ for a test suite $T$ is a sequence $(t_1,\ldots,t_m)$ as a permutation of the elements in $T$. We should note that the test suites may evolve over time, and it is possible to add or remove tests in different software revisions. A schedule fitness metric is a function $f: S \to \mathbb{R}$ that produces a score for the given schedule.

In this article, we will focus on Average Percentage of Fault Detection (APFD). A test case prioritization procedure is any function that given a set of tests $T$ produces a test suite as a sequence $(t_1,\ldots,t_m)$. Ideally, the schedule should be optimal with respect to a given fitness function. With this objective, TCP algorithms often use some kind of  test case failure estimator, since most fitness functions favor the execution of failing tests first. A failure estimator is a function that, given a test case $t$ and historic information about the software revisions $S$ as well as its development, provides an estimate of the probability that the outcome of the test is a failure for the current software revision.

Our solution to the TCP problem is implemented by combining the application of a static scheduler and a dynamic one. First, all tests are scheduled with the static scheduler. After this, the test with the highest score as ranked by the static scheduler is executed. The verdict of this test is then used by the dynamic scheduler to reschedule the pending tests. The next test is again selected for execution, and once more, its verdict is used to reschedule the pending tests. This process is repeated until all tests are processed.

The advantage of this approach is that different schedulers can be combined in a single TCP algorithm. As a static scheduler we can use any method that assigns priority scores. We propose a novel algorithm for dynamic scheduling in Section~\ref{sec-dynamic-with-condit-prob}. This algorithm is based on the conditional probability for pairs of tests to have the same verdict in a test cycle, and is independent of the static scheduler.

\section{Dynamic Scheduling using Conditional Probability}
\label{sec-dynamic-with-condit-prob}
This section presents the dynamic scheduler we use\cmmnt{in combination with the algorithm presented in section~\ref{sec-static-and-dynamic-tcp}}. The main idea of the algorithm is that it not only uses a test case failure estimator based on previous revisions $S$ to create a schedule $T$, but also applies knowledge gained from the current execution. First, each test case $t$ receives an individual score based on the ranking assigned by the static scheduler, according to the initial order created by this scheduler. Second, the algorithm analyzes which test cases tend to fail or pass together, again based on previous revisions. We achieve this by searching for pairwise correlations between tests. By pairwise correlations we mean that two test cases fail or pass together in a given software revision $S$. Cho et al. used a related approach, in which flips were investigated, ie, correlations between \emph{ changes} in verdict, \cite{ChoKL16afsac}. In our algorithm, the individual score of each test is updated based on the correlations and the verdicts of the executed tests.

The individual score of a test is inversely proportional to its scheduling rank, that is, the test scheduled at position $n$ will be assigned a score $\frac{1}{n}$. Pairwise correlations between tests are computed as conditional probabilities in the following ways: \begin{math} P(t_1=f | t_2=f)=\frac{P(t_1=f \cap t_2=f)}{P(t_2=f)} \end{math} and 
\begin{math} P(t_1=p | t_2=p)=\frac{P(t_1=p \cap t_2=p)}{P(t_2=p)} \end{math}. We remind the reader that conditional probability, is the likelihood of event B happening, given that we know that A has happened. To compute conditional probabilities, we chose a certain number of revisions and we will refer to this as history length (see subsection \ref{sec-westermo-dataset} for more details).

\begin{algorithm}
\caption{Dynamic Test Case Prioritization with conditional probability}\label{alg2:cap}

\textbf{Input:} \emph{t\_to\_execute, t\_scores, t, t\_verdict, t\_corr\_fail, t\_corr\_pass}
\textbf{Output:} \emph{t\_scores}

\If{$t\_verdict = f$} {
    \For{\texttt{t\_corr in t\_corr\_fail}} {
        \If{$t\_corr \in t\_to\_execute$} {
            $t\_score \gets t\_scores.t\_corr + k*t\_corr\_fail.t\_corr$\;
            $t\_scores \gets substitute (t\_corr,\ t\_score)$\;
        }
    }
}
\If{$t\_verdict = p$} {
    \For{\texttt{t\_corr in t\_corr\_pass}} {
        \If{$t\_corr \in t\_to\_execute$} {
            $t\_score \gets t\_scores.t\_corr - k*t\_corr\_pass.t\_corr$\;
            $t\_scores \gets substitute (t\_corr,\ t\_score)$\;
        }    
    }
}     
\Return
\end{algorithm}

The Dynamic Test Case Prioritization algorithm with conditional probability is shown as Algorithm \ref{alg2:cap}\cmmnt{and is an instantiation of the approach described in section \ref{sec-static-and-dynamic-tcp}}. Depending on the verdict of the currently executed test, the individual scores of the correlated tests are either increased or decreased. For these correlated tests, tests which tend to fail or pass together are considered, and the conditional probabilities of these events are computed before the actual scheduling (corresponding to $t\_corr\_fail$ and $t\_corr\_pass$ in Algorithm \ref{alg2:cap}), as shown earlier in this section. Individual test scores for these correlated tests (the variable $t\_scores.t\_corr$) are either increased or decreased by the constant $k$ multiplied by the conditional probability of failure or success accordingly. Then, when the individual scores of tests are updated based on the outcome, the dynamic scheduling algorithm chooses the test with the highest individual score assigned by a static scheduler, and again, when this test is executed, based on its verdict, the individual score of each of the correlated tests is either increased or decreased. The algorithm terminates when all tests have been executed.

As we initially evaluated our algorithm with the Westermo dataset \cite{strandberg2022dataset} and learned that $k = 0.8$ demonstrated good performance, we applied the same constant with Paint Control \cite{SpiekerGMM18retecs} and IOF/ROL \cite{SpiekerGMM18retecs}. Since, ideally, this value should be selected individually for each test system, the expected improvement for the algorithm is that the value for $k$ is dynamically adjusted.

\section{Evaluation}
\label{sec-evaluation}
To effectively evaluate the Dynamic Test Case Prioritization approach we compare our solution to Optimal, Worst, and Random algorithms. The Optimal algorithm creates an optimal schedule, i.e., it always schedules all failing tests before the passing ones, thus the dynamic algorithm can either reduce or provide the same performance as the Optimal algorithm. The Worst algorithm acts in the opposite way, and the dynamic scheduler can only improve or provide the same performance as the Worst algorithm. The Random algorithm, as the name reveals, creates a schedule in a completely random manner. Due to non-deterministic nature of the Random algorithm, we execute it 30 times for each evaluated cycle. The dynamic approach can reduce, improve, or provide the same performance as the Random algorithm.

The solution is developed in Python with the help of  libraries such  as numpy and pandas. In the evaluation, we apply the APFD metric after each software revision.\cmmnt{Usually, a test case can reveal one or more faults, and a fault can be revealed by several tests. The APFD metric can distinguish between different faults, because it considers only the first test that reveals that fault.} In this article, we assume that each individual test case reveals one unique fault, because we use only test verdicts in our solution. The Python library called dnn-tip (version 0.1.1) \footnote{The library is available at https://pypi.org/project/dnn-tip/.} was utilized to compute APFD. This library was developed for \cite{WeissT22nntcp} and then released.

\subsection{General description of the datasets}
Our code was originally developed for an AIDOaRt\footnote{AIDOaRt is a European project focused on AI-augmented automation for Cyber-Physical Systems, see more at https://www.aidoart.eu/.} Hackathon and we evaluated our solution on a dataset provided by Westermo. Then, to evaluate our solution on more industrial datasets, we modified our code to adjust it to different data formats. As additional datasets we chose Paint Control and IOF/ROL which were created by ABB Robotics Norway \cite{SpiekerGMM18retecs}. The general overview of these datasets is shown in Table \ref{table:datasets}.

\begin{table}[h!]
\centering
\begin{tabular}{lrrcr}
 \hline
 Dataset      & Tests & Cycles & Verdicts        & Failed \\  \hline
 Westermo 11   &  468  &     92 & 29,815 (30,135) &   7.4 (7.8)\% \\
 Westermo 13   &  136  &    451 & 40,425 (40,426) & 11.9\% \\
 Westermo 21   &  267  &    204 & 28,349 (28,552) & 10.2 (10.1)\% \\
 Westermo 28   &  344  &    261 & 37,925 (37,936) & 8.3\% \\
 Westermo 30   &  335  &    445 & 86,834 (86,847) & 8.9\% \\
 Westermo 33   &  298  &    426 & 75,620 (75,623) & 9.6\% \\
 Westermo 46   &  329  &    415 & 81,141 (81,299) & 5.5 (5.4)\% \\
 Westermo 64   &  347  &    411 & 76,081 (76,109) & 9.7 (9.8)\% \\
 Westermo 96   &  310  &    169 & 30,574 (30,583) & 14.7\% \\
 Paint Control &   89  &    352 & 22,260 (25,594) & 15.2 (19.4)\%  \\
 IOF/ROL       & 1941  &    320 & 27,664 (32,260) & 17.9 (28.8)\%  \\
 \hline
\end{tabular} 
\caption{Characteristics of datasets. A similar approach is presented in \cite{PradhanWAYL18remap, PradhanWAYL19multiobjectivesearch, SpiekerGMM18retecs} for Paint Control and IOF/ROL.}
\label{table:datasets}
\end{table}
As shown in Table \ref{table:datasets}, the Westermo dataset consists of nine separate test systems\footnote{The chosen setting: results from nightly testing performed on the main branch, since this type of testing was the regular one at Westermo. See \cite{strandberg2022dataset} for more details on the dataset.}, while Paint Control and IOF/ROL include one test system each. For each dataset or each test system, we computed the number of unique test cases, the number of test cycles, the number of verdics, which is a total number of tests executed in all test cycles, as well as the percentage of failed test cases.  As mentioned in \cite{PradhanWAYL18remap} and \cite{PradhanWAYL19multiobjectivesearch} as well as through our discussions with Westermo, we learned that sometimes the same test case can be executed several times in a CI cycle and the failure of this test case may not be caused by an error in the system under test. Because of this, we pre-processed the data before executing our algorithm and left only the last verdict of each test case that was executed several times in a test cycle. In Table \ref{table:datasets}, the values in parentheses relate to statistics from the original datasets, while the values without parentheses belong to the statistics from the modified datasets. If there is only one value shown in the cell, it means that this value stayed unchanged.\cmmnt{For percentage of failed test cases, we rounded values to one decimal place to adjust to the layout.}\cmmnt{, thus two values are present in one cell if these differ after being rounded. In total, the number of verdicts in all 11 test systems is 536,688 after the modification.}

\subsection{The Westermo dataset} 
\label{sec-westermo-dataset}
Among the historical data provided in the Westermo dataset \cite{strandberg2022dataset}, features that were utilized for the current algorithms are verdicts of test cases. We treat each of the nine test systems (see Table \ref{table:datasets}) separately, as each of these creates a unique environment. \cmmnt{In the dataset, test suites can be distinguished based on the test system (see numerical values in Table \ref{table:datasets}) and type of testing along with the test branch. From the description of the Westermo dataset, each test system corresponds to a certain unique set of configurations required to perform testing, and the type of testing means a certain purpose: a regular testing activity performed during nights, testing of software with reported issues, and testing of new or suspicious tests. Among the provided testing types, as already mentioned, we considered nightly testing performed on the main branch, and each of nine test systems was treated separately, as each of these creates a unique environment. Even if the same test can be run on several test systems, its verdicts may differ depending on which system they were applied to. The method currently in use applies a heuristic to sort tests in order of priority based on a few chosen properties\cite{strandberg2016experience} and is static.} As each test system in the dataset includes a large number of revisions, we chose a range of latest 300 revisions for the evaluation. For systems that contain less than 300 cycles, we selected fewer cycles for the evaluation considering the chosen history length. By history length we mean how many previous revisions would be considered for prediction of correlations. In several publications, for example in \cite{ShiXW20rlfortcp} and \cite{SpiekerGMM18retecs}, the authors used the history length of four as it was empirically proven that it leads to a better performance of reinforcement learning-based techniques. For our algorithms, we observed that too short a history window led to worse performance. The required history length differs between systems, thus an individual parameter should be chosen for each of these. In the current experiments, to apply the same parameter to all systems, the history length was set to 15, since we observed that this history length provided a better performance for the majority of test systems. The expected improvement for the algorithm is that the value of the history length is adjusted for each test system dynamically. Since our goal is to evaluate how the dynamic approach can influence the performance of a static one, we consider only test cycles where there are both failing and passing tests. By fault we consider test verdicts $fail(1)$ and $invalid(2)$, while $0$ corresponds to the passed test. In addition to that, the dataset contains test verdicts equal to $3$ which means that some resources were not available when the test should have been executed, but we do not consider these verdicts neither for correlation scores nor for evaluation of our algorithm.



\subsection{The Paint Control and the IOF/ROL datasets}
Paint Control and IOF/ROL are industrial datasets created by ABB Robotics Norway and contain data from testing of robots on a daily basis \cite{SpiekerGMM18retecs}. For both datasets, the approach chosen for range of cycles and history length is the same as for the Westermo dataset. Unlike Westermo, the verdicts of test cases for Paint Control and IOF/ROL are divided only to two groups: 0 if the test passed and 1 if it failed. Unfortunately, we do not have as much information about these datasets as in the Westermo dataset, thus we can rely only on the information we can derive from the datasets themselves. 

\subsection{Results}
\begin{figure}[t]
    \centering
    \begin{tabular}{lll}
        \multicolumn{1}{c}{Static: Optimal} & 
        \multicolumn{1}{c}{Static: Random} & 
        \multicolumn{1}{c}{Static: Worst} \\
        \FPFC{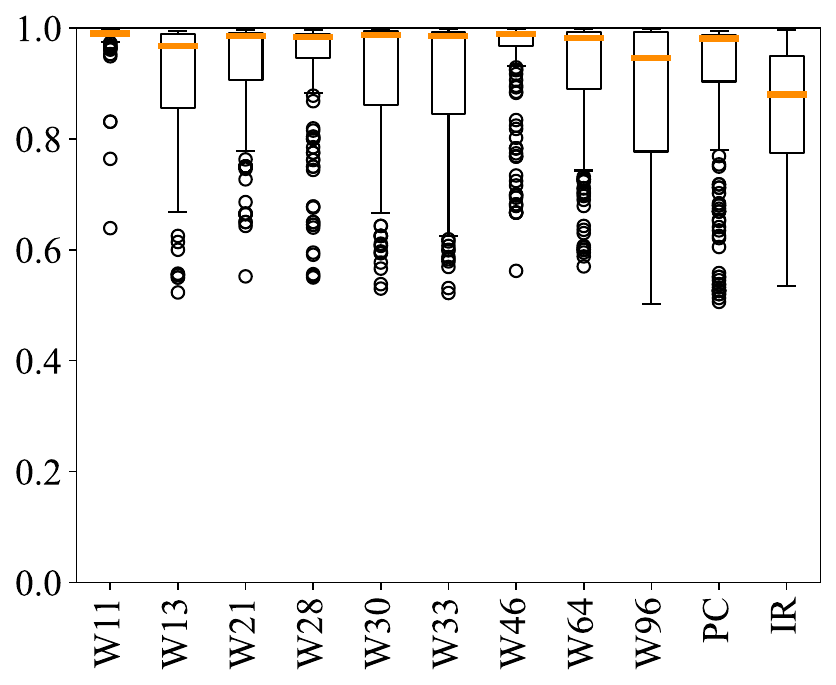}& \FPFC{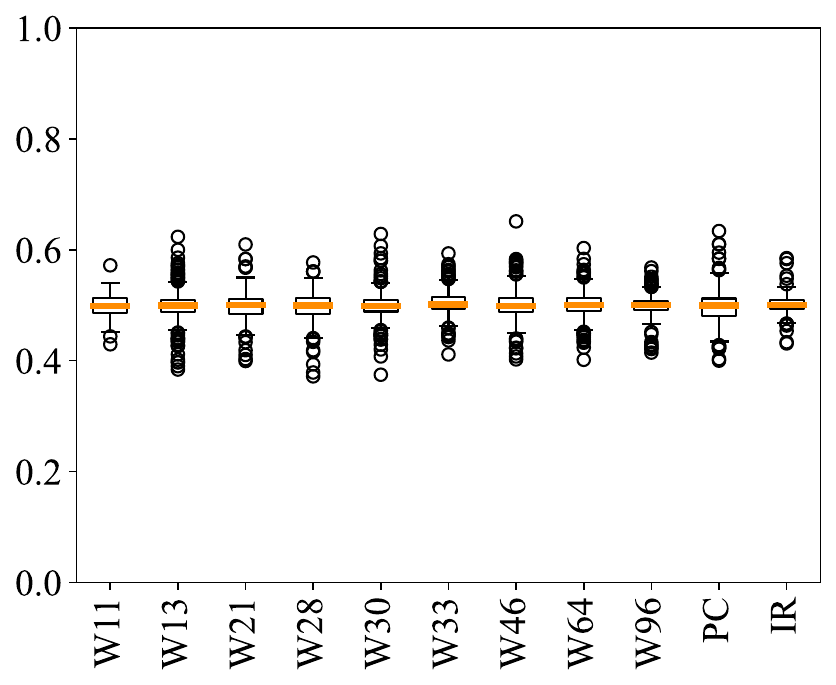}& \FPFC{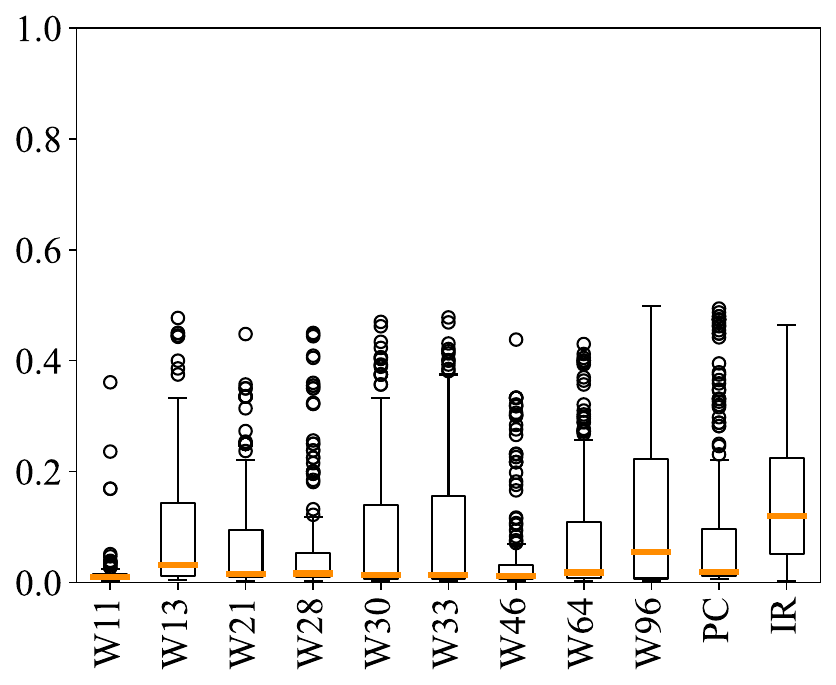}\\
          \multicolumn{1}{c}{+ dynamic: CP} &
          \multicolumn{1}{c}{+ dynamic: CP} &
          \multicolumn{1}{c}{+ dynamic: CP}\\
         \FPFC{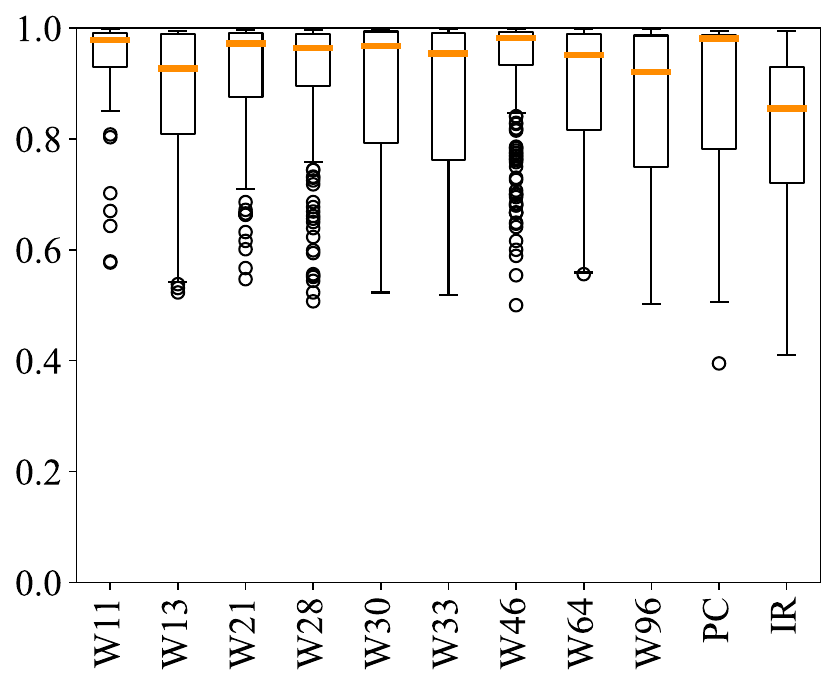} & 
         \FPFC{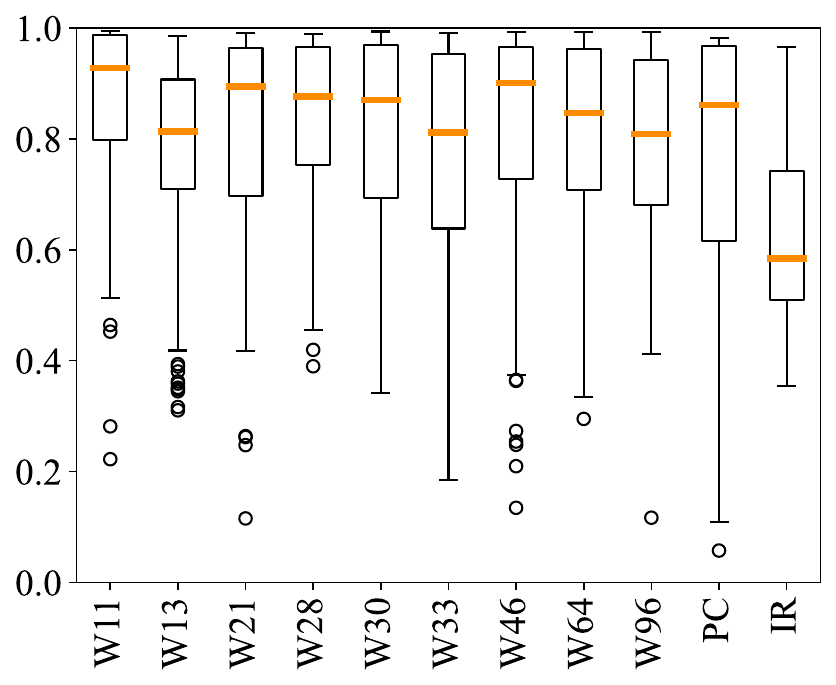}  &
         \FPFC{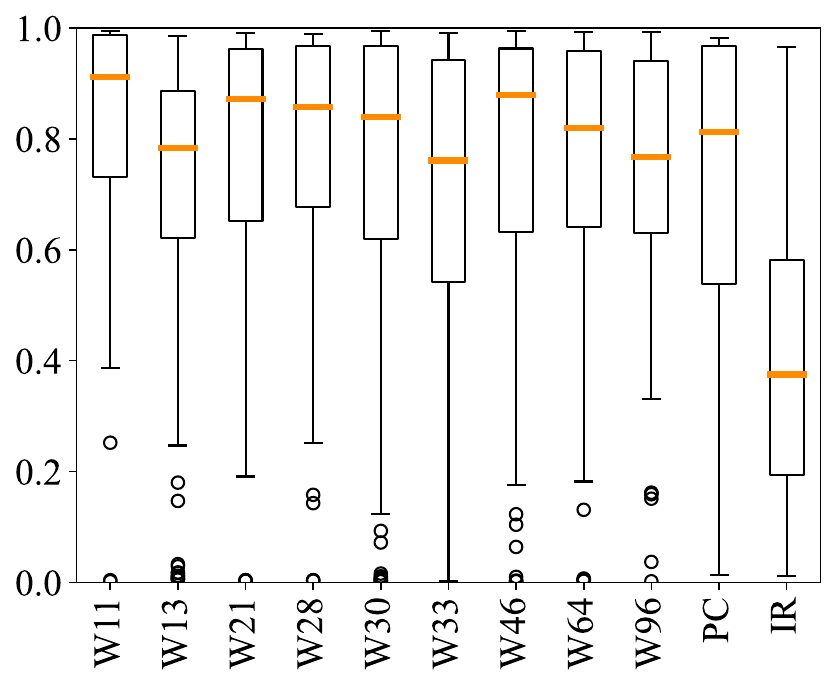} \\
    \end{tabular}
    \caption{Comparison for APFD for static and dynamic prioritization algorithms (higher values are better). The first nine systems are from the Westermo dataset, the last two are from Paint Control and IOF/ROL.}
    \label{tab:apfd}
\end{figure}

We compare the three mentioned static algorithms to the following dynamic ones: Optimal + Conditional Probability, Random + Conditional Probability, and Worst + Conditional Probability. Figure \ref{tab:apfd} demonstrates evaluation of performance in terms of APFD for Westermo, Paint Control, as well as IOF/ROL datatasets. The static approach for Optimal, Random, and Worst algorithms is compared with the dynamic one. As random algorithms were executed 30 times for each cycle, the mean APFD for each cycle is considered. The APFD metric demonstrates a slight decrease in the performance of the dynamic approach when applied to the Optimal algorithm, however in practice it is not a big drawback since practitioners rarely have optimal test suites. Among mean values, the dynamic approach improves the performance for both Random and Worst algorithms, while the most visible change can be seen for the Worst algorithm, where for the majority of cases median values increase with more than 0.8 when comparing to these values for the static approach. The dynamic approach applied to the Random and Worst algorithms showed the lowest improvement for the IOF/ROL dataset. This may mean that tests in this dataset are not correlated.

\section{Discussion and Conclusions}
\label{sec-conclusions}
In this article, we presented a novel dynamic test case prioritization algorithm that can be applied with any static prioritizer that assigns a priority score to each test case. We use such static algorithms as Worst, Random, and Optimal to evaluate how the dynamic prioritization algorithm can improve the schedules created by these. Our algorithm relies on scores assigned by the chosen static algorithm, as well as correlation between test cases derived from the history of the chosen time interval. These correlations help to change an individual score of related pending tests and thus reschedule these tests. We evaluated our approach with Westermo, Paint Control, and IOF/ROL datasets. If there are correlations between test cases, the dynamic approach helps to improve the schedule so that faults are revealed faster, both for the Random and for the Worst static algorithms, thus conditional test failure probability helps to improve the schedule created by a static algorithm. If the static schedule is optimal, the dynamic schedule may decrease the performance, which is expected. Thus,\textit{ dynamic test case prioritization improves suboptimal test schedules, in most cases the improvement is substantial}.

For the industry, our approach helps to dynamically adjust schedule of test cases and execute possibly failing test cases earlier, thus decreasing the waiting time for developers. When evaluated with the APFD metric, the dynamic algorithm demonstrated a noticeable increase in performance when applied to Random and Worst algorithms which means that it succeeded to schedule failing test cases earlier. The limitations of our approach are that constant values are applied to choose history length as well as to adjust scores of correlated test cases. As a possible direction for the future work, these values should be adjusted dynamically for each test system. In addition to that, the dynamic approach should be applied to existing static TCP baselines.



\begin{acks}
This research has received funding from the ECSEL Joint Undertaking (JU) under grant agreement No 101007350. The JU receives support from the European Union’s Horizon 2021 research and innovation program and Sweden, Austria, Czech Republic, Finland, France, Italy, Spain.
\end{acks}




\bibliographystyle{plain}
\balance

\end{document}